\documentclass[a4paper]{article}

\usepackage{INTERSPEECH2021}
\newcommand\norm[1]{\left\lVert#1\right\rVert}
\usepackage{algorithmic}
\usepackage{soul, color}
\usepackage{xcolor}
\usepackage[linesnumbered,ruled,vlined,noend]{algorithm2e}

\SetCommentSty{mycommfont}
\SetKwInput{KwInput}{Input}                
\SetKwInput{KwOutput}{Output}              
\SetKwInput{KwRequire}{Require}              

\usepackage[caption=true,font=footnotesize]{subfig}

\title{FastICARL: Fast Incremental Classifier and Representation Learning with Efficient Budget Allocation in Audio Sensing Applications}
\name{Young D. Kwon$^1$, Jagmohan Chauhan$^{1,2}$, Cecilia Mascolo$^1$}
\address{
  $^1$University of Cambridge, UK\\
  $^2$University of Southampton, UK}
\email{\{ydk21,cm542\}@cam.ac.uk, j.chauhan@soton.ac.uk}

\begin{document}

\maketitle

\begin{abstract}

    Various incremental learning (IL) approaches have been proposed to help deep learning models learn new tasks/classes continuously without forgetting what was learned previously (i.e., avoid catastrophic forgetting). With the growing number of deployed audio sensing applications that need to dynamically incorporate new tasks and changing input distribution from users, the ability of IL on-device becomes essential for both efficiency and user privacy. 
    

    However, prior works suffer from high computational costs and storage demands which hinders the deployment of IL on-device. In this work, to overcome these limitations, we develop an end-to-end and on-device IL framework, FastICARL, that incorporates an exemplar-based IL and quantization in the context of audio-based applications. 
    We first employ k-nearest-neighbor to reduce the latency of IL. Then, we jointly utilize a quantization technique to decrease the storage requirements of IL. We implement FastICARL on two types of mobile devices and demonstrate that FastICARL remarkably decreases the IL time up to 78-92\% and the storage requirements by 2-4 times without sacrificing its performance. FastICARL enables complete on-device IL, ensuring user privacy as the user data does not need to leave the device.
    

\end{abstract}

\section{Introduction}\label{sec:Introduction}

A recent development of deep learning has revolutionized various audio-based applications such as emotion recognition (ER)~\cite{rachuri_emotionsense_2010}, environmental sound classification (ESC)~\cite{salamon_dataset_2014}, and keyword spotting~\cite{mathur_unsupervised_2020,mittal_representation_2020}. However, in a real-world setting where a deployed audio classification models may need to dynamically incorporate new tasks (i.e., new classes or inputs) from users~\cite{chauhan_contauth_2020} and changing input distribution~\cite{pan_survey_2010}, current supervised learning approaches are severely limited due to the constrained nature of available resources on the edge devices and the catastrophic forgetting (CF) issue~\cite{mccloskey_catastrophic_1989}. That is, a deep learning model becomes able to recognize a new task but forgets previously learned knowledge.

Many researchers proposed a range of Incremental Learning (IL) methods~\cite{parisi_continual_2019} to solve the CF problem. The first group of the IL approaches is a \textit{regularization-based method}~\cite{kirkpatrick_overcoming_2017,zenke_continual_2017,schwarz2018progress} where regularization terms are added to the loss function to minimize changes to important weights of a model for previous tasks to prevent forgetting. Kirkpatrick et al.~\cite{kirkpatrick_overcoming_2017} proposed a regularization-based method, Elastic Weight Consolidation (EWC), which uses the Fisher information matrix to identify important weights to the previous tasks and update less on those weights while learning a new task. Another group of the IL approaches is \textit{exemplars-based method}~\cite{rebuffi_ICARL:_2017,lopez-paz_gradient_2017} where the method requires to store important samples from previous tasks to prevent from forgetting learned tasks. Rebuffi et al.~\cite{rebuffi_ICARL:_2017} proposed a representative exemplar-based method, ICARL, that first utilizes herding~\cite{welling_herding_2009} to search for exemplars (informative samples) and then uses knowledge distillation loss on the previously learned classes and classification losses on a new class to prevent forgetting and learn the new class. However, prior works are limited in two ways. First, It is challenging to enable IL on-device since IL methods are computationally heavy. Second, exemplar-based methods require storing exemplars, which can impose a considerable burden on resource-constrained systems.

Moreover, many techniques have been proposed to facilitate efficient machine learning systems on resource-constrained devices. Quantization and low-bit precision of model parameters are utilized to reduce the size of the model~\cite{jacob_quantization_2018, courbariaux_binarized_2016}.  Low-rank factorization~\cite{lane_deepx_2016,Pang2018} and pruning~\cite{han_deep_2016} have been proven effective in reducing model size, while retaining accuracy. 
IL with optimizations that allow its use on-device, however, has never been explored in the context of audio-based applications.

In this work, an end-to-end framework, FastICARL, is developed to enable efficient and accurate on-device IL in two audio sensing applications, an ER task and an ESC task. 
Also, FastICARL is a new IL method devised to improve upon the representative exemplar-based IL method, ICARL, as we observed that ICARL consistently outperforms EWC and other regularization-based IL methods~\cite{zenke_continual_2017,schwarz2018progress}. However, it has computational and storage issues. Thus, FastICARL solves these limitations while maintaining accuracy. First, we optimize the construction process of an exemplar set (which takes most of the IL time) to shorten the IL time to tackle the first limitation. 
Specifically, to find the informative exemplars that can best approximate feature vectors over all training examples, ICARL relies on herding which contains inefficient double for loops. Instead, FastICARL utilizes a k-nearest-neighbor and a max heap data structure to search exemplars more efficiently. In addition, to address the second limitation, we further optimize FastICARL by applying quantization on exemplars to reduce the storage requirement. 
We convert the 32-bit float data type into 16-bit float and 8-bit integer data types. Furthermore, we implement our end-to-end IL framework on mobile and embedded devices of two different specifications: Jetson Nano and a smartphone (Google Pixel 4). 
For a smartphone implementation, we employ MNN~\cite{jiang_mnn_2020} and our implementation enables complete on-device training of new tasks/classes unlike TensorFlow Lite~\cite{lee_-device_2019} or PyTorch Mobile~\cite{noauthor_pytorch_nodate} where only on-device inference is enabled.

Overall, the major contributions and findings of this paper are as follows. We design, implement, and evaluate FastICARL, which overcomes the limitations of the prior work. First of all, FastICARL shows that it can effectively solve the CF issues happening in audio-based datasets by achieving 69\% and 71\% weighted F1-scores for ER and ESC, respectively. FastICARL reduces the latency of exemplar set selection up to 78\% on Jetson Nano and 92\% on Google Pixel 4. Moreover, FastICARL decreases the storage requirement by 2-4 times without sacrificing its performance. In addition, we demonstrate that FastICARL can enable on-device IL without the support of the cloud. Hence, FastICARL ensures complete data privacy as user data does not need to leave the device. Finally, to the best of our knowledge, FastICARL is the first end-to-end and on-device framework that incorporates exemplar-based IL and quantization techniques in the context of audio sensing applications.






\section{Methodology}\label{sec:Methodology}
In this section, we formulate our problem (\S\ref{subsec:Problem Formulation}) and describe the important prior work (\S\ref{subsec:ICARL}). After that, we propose our IL method, FastICARL (\S\ref{subsec:FastICARL}).

\subsection{Problem Formulation} \label{subsec:Problem Formulation}


We focus on Sequential Learning Tasks (SLTs)~\cite{pfulb_comprehensive_2018} from the audio sensing tasks,
where new classes (e.g., different sounds in ESC) can emerge over time. Thus, the learning model has to continuously learn to accommodate new classes without CF, as would happen in real-life scenarios. Learning tasks of this type, called SLTs, indicates that a model continuously learns two or more tasks $D_1,..., D_t$, one after another instead of learning a single task $D$ once (i.e., multi-task learning). Note that each task consists of disjoint groups of classes as we adopt class-incremental learning~\cite{van_de_ven_three_2019}. Formally, at any time, we are given training samples, $X^1, X^2,...,X^N$, where $X^y$ is a set of samples of class $y$ and $N$ is the total number of given classes. Inspired by prior works~\cite{kemker_measuring_2017, chauhan_contauth_2020}, we first train a model on the first task with $N/2$ classes and then incrementally train the model by adding subsequent tasks with one class. In total, $1+N/2$ tasks are learned incrementally.

\subsection{ICARL}\label{subsec:ICARL}
ICARL is the representative exemplar-based IL method in the literature that attempts to solve the CF problem of class-incremental setting. At the high level, ICARL maintains a set of exemplar samples for each observed class (see Algorithm~\ref{alg:exemplar}). An exemplar set is a subset of all samples of the class to carry the most representative information of the class. When new tasks (classes) become available, ICARL first creates a new training set by joining all exemplar sets and the data of the new class. Then, it updates its weight parameters by minimizing a classification loss of the new task (class) as well as the distillation loss of the previous tasks (classes). 
Then, ICARL builds an exemplar set for the new class and trims the existing exemplars for previous classes. Finally, the classification is performed by finding the nearest-class-mean of exemplars to a given test sample in a feature space extracted from the learned representation.


\begin{algorithm}[!t]
\caption{Construction and quantization of exemplar sets for ICARL/FastICARL}
\label{alg:exemplar}
\SetAlgoNoLine
\DontPrintSemicolon
  \KwInput{Feature Extractor $\mathcal{F()}$, The number of exemplars to be stored $m$, Quantization bit $b$, IL method}
  \KwOutput{Quantized Exemplar set $Q$}
  \KwData{$X = \{x_1, ..., x_n\}$ of class $y$}
    
    $\mu \leftarrow \frac{1}{n} \sum_{i=1}^{n} \mathcal{F}(x_i) $ \tcp*{calculate class mean}
    
    \tcc{find m exemplars out of n samples}
    \If{ \textnormal{IL method is} \textit{ICARL}}
    {
        \For{$k = 1, ..., m$}
        {
            $p_k \leftarrow \underset{x \in X}{\mathrm{argmin}}\norm{\mu - \frac{1}{k} ( \mathcal{F}(x) + \sum_{i=1}^{k-1} \mathcal{F}(p_i) ) }$\;
        }
    }
    
    \If{ \textnormal{IL method is} \textit{FastICARL}}
    {
        \tcc{calculate feature distance between each sample and class mean}
        \For{$i = 1,...,n$}{
            $d_i = \mathcal{F}(x_i) - \mu$
        }
        
        \tcc{build max heap with size m}
        create max heap $H$ of pair \{d, index\}\;
        
        \For{$k = 1, ..., m$}{
            H.insert( $d_k, k$ )\;
        }
        \tcc{loop over the remaining samples while updating the max heap}
        \For{$k = m+1, ..., n$}{
            \If{$d_k < H.\textnormal{extractMaxDist}()$}{
                $H$.pop() \tcp*{delete one item from H}
                $H$.insert( $d_k, k$ )\;
            }
        }
        
        \tcc{build a sorted exemplar set P}
        \For{$k = m, ..., 1$}{
            $i \leftarrow H$.extractMaxDistIndex(), $H$.pop()\;
            $p_k \leftarrow x_i$\;
        }
        
    }
    
    \For{$k = 1, ..., m$}
    {
        $q_k \leftarrow \textnormal{Quantize}(p_k, b)$\;
    }
    $Q \leftarrow (q_1, ..., q_m)$ \tcp*{Quantized exemplar set}
    
\vspace{-0.1cm}
\end{algorithm}

\subsection{FastICARL}\label{subsec:FastICARL}
%
Although ICARL provides impressive performance,
it is limited by high computational costs and large storage requirements to maintain sufficient budget size to perform reasonably well.
To begin with, ICARL's high computational loads comes from its herding operation (find an exemplar set that has a min distance between the class mean and exemplars mean in feature space), i.e., exemplar selection procedure which is based on the inefficient double for loops (Lines 2-4), resulting in the $O(nm^2)$ complexity (which takes up 70 - 90\% of the total IL time). $n$ is the number of examples in a class, and $m$ represents the target number of exemplars. 
Note that in this work, training time indicates the usual training time with respect to back-propagation, updating weights, while the rest of the time in learning a new task or adding a new class is considered IL time.
Thus, instead of relying on herding, FastICARL employs a k-nearest-neighbor search to identify the representative examples to construct exemplar sets. 
This enables FastICARL to accelerate the process of exemplar construction without performance degradation, as shown in Section~\ref{sec:Evaluation}.
By jointly utilizing the max heap as in Algorithm~\ref{alg:exemplar}, FastICARL remarkably reduces the complexity of finding $m$ exemplars out of $n$ samples to $O(n(1+log(m)) + mlog(m)) = O(nlog(m))$. In detail, the computation of feature distance and the insertion of max heap cost $1+log(m)$ which is performed on $n$ samples in total. After that, the sorting on $m$ identified exemplars in a max heap costs another $mlog(m)$.


Furthermore, ICARL requires as much as 69 MB (see \S\ref{subsec:Results}). To alleviate this storage demand, we apply quantization on exemplar sets on the fly. Note that since budget sizes take up 72-99\% of the storage requirements of FastICARL, we apply quantization only on exemplars in this work. While constructing exemplar sets, FastICARL converts 32-bit float data to 16-bit float or 8-bit integer types and store them with a smaller budget. When converting between 32-bit float and 8-bit integer, we use quantization scheme used in~\cite{jacob_quantization_2018} to minimize the information loss in quantization. We utilize an affine mapping of integer q to real number r
for constant quantization parameters $S$ and $Z$, i.e.,
$r = S (q - Z)$. $S$ denotes the scale of an arbitrary positive real number. $Z$ denotes zero-point of the same type as quantized value q, corresponding to the real value 0.

\section{Evaluation}\label{sec:Evaluation}

\subsection{Datasets}\label{subsec:Datasets}
We experiment with our method on two audio applications.

\noindent\textbf{EmotionSense:} For emotion recognition (ER) application, we employ the EmotionSense dataset~\cite{rachuri_emotionsense_2010} as it is used in multiple studies in audio sensing~\cite{georgiev2014dsp, lane_deepear:_2015, servia-rodriguez_knowing_2021}. It contains audio signals which are clustered into five standard broader emotion groups, generally used by social psychologists~\cite{feldman1998independence} such as (1) Happy, (2) Sad, (3) Fear, (4) Anger, and (5) Neutral. This dataset has 2,235 samples, and each measurement corresponds to a particular emotion based on a 5-second context window. Following~\cite{georgiev_low-resource_2017}, we extract 24 log filter banks~\cite{smith_new_1987} from each audio frame over a time window of 30 ms with 10 ms stride. After that, as our preprocessing steps, we downsample each sample measurement by averaging corresponding 24 filter banks of every 250 ms (or 25 consecutive windows) without any overlap to reduce the length of the input sequence for a learned neural network. We normalize each window to zero mean and unit variance. As a result, we created an input of size $20 \times 24$.

\noindent\textbf{UrbanSound8K:} For environment sound classification (ESC) application, we adopt the UrbanSound8K dataset~\cite{salamon_dataset_2014} as it is a large dataset that can test the effectiveness of our method on resource-limited devices. UrbanSound8K contains 9.7 hour-long data with 8,732 labeled urban sounds collected in real-world settings. This dataset consists of 10 audio event classes such as car horn, drilling, street music, etc. Following~\cite{su_environment_2019}, we extracted four different audio features ((1) Log-mel spectrogram, (2) chroma, (3) tonnets, (4) spectral contrast) for each sound clip, sampled at 22 kHz. Using the first 3-seconds of sound, we created an input of size $128 \times 85$, where 128 represents the number of frames and 85 represents aggregated feature size of the four audio features.

\subsection{Experimental Setup}\label{subsec:Experimental Setup}

\textbf{Task:} As described in \S\ref{subsec:Problem Formulation}, we adopt class-incremental learning. Hence, for EmotionSense, two classes are selected as task 1 for training a base model, and then the other three classes are added to the model one by one sequentially. For UrbanSound8K, five classes are used as the first task, and the other five classes are learned incrementally. Note that all reported results in \S\ref{subsec:Results} are averaged over five times of experiments.

\noindent\textbf{Model Architecture:} We adopt a convolutional neural networks (CNN) architecture from prior work~\cite{su_environment_2019} to construct the ER and ESC models. To identify a high-performing and yet lightweight CNN model to operate on embedded and mobile devices, we conducted hyper-parameter search with different number of convolutional layers \{2,3,4\}, number of convolutional filters \{8,16,32\}, pooling layer type \{max pooling, average pooling\}, number of fully-connected (FC) layers \{0,1\} and its hidden units \{128,512,1024\}. A basic convolutional layer consists of $3 \times 3$ convolution, batch normalization, and Rectified Linear Unit (ReLU). 
We found that although the best performing model is a 4-layered CNN with 32 Conv filters followed by an FC layer (Weighted F1-score of 86\% for ER and 90\% for ESC), the performance degradation without the FC layer is minimal (see Table~\ref{tab:performance}) while the majority of the model parameters are consumed in the FC layer as shown in~\cite{su_environment_2019}. 
Hence, as our final CNN architecture, we use [Conv: \{32,32,64,64\}] for ER and [Conv: \{16,16,32,32\}] for ESC. We omit an FC layer in both applications, and average pooling layers and a 0.5 dropout probability are adopted for the second and fourth Conv layers. ADAM optimizer~\cite{kingma_adam_2017} and learning rate of 0.001 are used.

\noindent\textbf{Evaluation Protocol:} 
Following prior works~\cite{rachuri_emotionsense_2010, su_environment_2019}, the 10\% of each class is used as the test set and the remaining as the training data. 
In addition, we report the performance of a model trained up to task k incrementally. Also, we report the results based on a weighted F1-score which is more resilient to class imbalances as the employed datasets are not balanced.

\noindent\textbf{Baselines:} To evaluate the effectiveness of FastICARL, we include various baselines in our experiments. First, we include a \textit{Joint} model which represents a scenario when the model is trained with training data of all classes available from the beginning. \textit{Joint} serves as a performance upper bound. Second, a \textit{None} model represents a case where a model is fine-tuned incrementally by adding classes to the model without any IL method. \textit{None} can be regarded as a performance lower bound. Thirdly, we include ICARL with three quantization levels (32, 16, and 8 bits). Finally, FastICARL (32, 16, and 8 bits) is compared.

\begin{table}[t]
  \centering
  \caption{
  Average weighted F1-score of baselines and FastICARL according to the budget size ($\mathcal{B}=5\%,10\%,20\%$) in EmotionSense and UrbanSound8K datasets.
  }
  \label{tab:performance}
  \resizebox{0.93\columnwidth}{!}{
  \begin{tabular}{ c | c c c | c c c }
    \toprule 
     &  \multicolumn{3}{c}{\textbf{EmotionSense (ER)}} & \multicolumn{3}{c}{\textbf{UrbanSound8K (ESC)}}\\
        \cmidrule(l){2-7}
     & 5\% & 10\% & 20\% & 5\% & 10\% & 20\% \\
        \cmidrule(l){1-7}
    ICARL (32 bits) & 0.57 & 0.60 & 0.70 & 0.67 & 0.69 & 0.69 \\
    ICARL (16 bits) & 0.55 & 0.63 & 0.70 & 0.66 & 0.67 & 0.71 \\
    ICARL (8 bits) & 0.59 & 0.62 & 0.68 & 0.65 & 0.68 & 0.70 \\
        \cmidrule(l){1-7}
    FastICARL (32 bits) & 0.57 & 0.62 & 0.67 & 0.67 & 0.69 & 0.70 \\
    FastICARL (16 bits) & 0.58 & 0.65 & 0.68 & 0.66 & 0.69 & 0.71 \\
    FastICARL (8 bits) & \textbf{0.60} & \textbf{0.63} & \textbf{0.69} & \textbf{0.65} & \textbf{0.68} & \textbf{0.69} \\
        \cmidrule(l){1-7}
    Joint (Upper Bound) & & 0.83 & & & 0.89 & \\
    None (Lower Bound) & & 0.41 & & & 0.02 & \\
        \bottomrule
  \end{tabular}
  }
  \vspace{-0.3cm}
\end{table}

\subsection{Implementation}\label{subsec:Implementation}

To evaluate our framework on resource-constrained devices, we implemented it on an embedded (Jetson Nano) and a mobile device (Google Pixel 4). The Jetson Nano is an embedded mobile platform with four cores and 4 GB RAM. It is often utilized in mobile robotics. 
We use PyTorch 1.6 to develop and evaluate FastICARL on Jetson Nano.
The Google Pixel 4 phone has eight cores and 6 GB RAM. 
We develop FastICARL based on C++ on the Android smartphone using mobile deep learning framework, MNN, and the Android Native Development Kit. 
Note that our implementation of FastICARL on the smartphone enables complete on-device training of new tasks/classes incrementally, unlike other deep learning frameworks on mobile platforms (e.g., PyTorch Mobile) where only on-device inference is supported.
The binary size of our implementation on a mobile platform is only 3.8 MB which drastically reduces the burden of integrating the IL functionality into mobile applications given that ICARL requires as much as 69 MB for UrbanSound8K.

\begin{table*}[t]
  \centering
  \caption{
  Average Latency (IL Time) in seconds for ICARL and FastICARL on Jetson Nano and a smartphone (Google Pixel 4) for both datasets according to the budget size ($\mathcal{B}=5\%,10\%,20\%$).
  }
  \label{tab:latency}
  \resizebox{1.77\columnwidth}{!}{
  \begin{tabular}{ c | c c c | c c c || c c c | c c c }
    \toprule 
     & \multicolumn{6}{c}{\textbf{Embedded Device (Jetson Nano)}} & \multicolumn{6}{c}{\textbf{Smartphone (Google Pixel 4)}}\\
        \cmidrule(l){2-13}
     & \multicolumn{3}{c}{\textbf{EmotionSense (ER)}} & \multicolumn{3}{c}{\textbf{UrbanSound8K (ESC)}} & \multicolumn{3}{c}{\textbf{EmotionSense (ER)}} & \multicolumn{3}{c}{\textbf{UrbanSound8K (ESC)}}\\
        \cmidrule(l){2-13}
     & 5\% & 10\% & 20\% & 5\% & 10\% & 20\% & 5\% & 10\% & 20\% & 5\% & 10\% & 20\% \\
        \cmidrule(l){1-13}
    ICARL (32 bits) & 6.25 & 7.24 & 9.35 & 102 & 144 & 271 & 1.41 & 1.98 & 2.73 & 41.5 & 75.5 & 138 \\
    ICARL (16 bits) & 6.30 & 7.40 & 9.25 & 100 & 144 & 270 & 1.48 & 1.99 & 2.74 & 44.6 & 78.8 & 139 \\
    ICARL (8 bits) & 6.27 & 7.40 & 9.25 & 120 & 178 & 292 & 1.43 & 1.99 & 3.04 & 45.4 & 77.7 & 146 \\
        \cmidrule(l){1-13}
    FastICARL (32 bits) & 5.10 & 5.18 & 5.18 & \textbf{60.6} & 60.8 & 60.7 & 0.88 & 0.90 & \textbf{0.83} & 10.5 & 10.8 & \textbf{10.4} \\
    FastICARL (16 bits) & \textbf{4.96} & 4.98 & 5.22 & 61.1 & 61.5 & \textbf{60.6} & 0.87 & 0.89 & 0.84 & 10.7 & 11.2 & 10.9 \\
    FastICARL (8 bits) & 5.01 & 5.07 & 5.24 & 67.1 & 66.3 & 61.5 & 0.90 & 0.91 & 0.87 & 10.7 & 10.7 & 10.6 \\
        \bottomrule
  \end{tabular}
  }
  \vspace{-0.3cm}
\end{table*}

\subsection{Results}\label{subsec:Results}


\noindent\textbf{Performance:} 
We first show the average weighted F1-score across all runs for different baselines and IL methods for the EmotionSense and UrbanSound8K datasets in Table~\ref{tab:performance}. For both datasets, we present the performance according to the size of the budgets storing exemplars (5\%, 10\%, and 20\%) to analyze trade-offs between the performance and storage requirement of the studied IL methods. Note that the weighted F1-score of the models after all tasks are trained incrementally is reported.

To begin with, the \textit{None} model allows us to confirm that CF occurs without the IL method. Its weighted F1-score drops sharply to 41\% for ER and 2\% for ESC. In contrast, the \textit{Joint} model achieves as high as 83\% and 89\% weighted F1-scores for ER and ESC, respectively.
ICARL (32 bits) and our proposed IL method, FastICARL (32 bits), can largely mitigate the CF issues observed in the None model. With a budget size of 20\%, ICARL provides a high weighted F1-score of 70\% for ER and 69\% for ESC. Likewise, FastICARL achieves a similar performance (67\% for ER and 70\% for ESC) to that of ICARL, which stays close to the upper bound performance of the \textit{Joint} model.
Furthermore, we find that the impact of the information loss due to the quantization of the saved exemplars for both ICARL and FastICARL is minimal (as observed in prior works about quantization~\cite{jacob_quantization_2018,han_deep_2016,krishnan_quantized_2019}).
As shown in Table~\ref{tab:performance}, all four variants, such as ICARL (16 and 8 bits) and FastICARL (16 and 8 bits), achieve similar performance to their original counterparts.

Finally, we study the importance of the storage budget parameter. We present the performance of our IL method according to its budgets of 5\%, 10\%, and 20\% of total training samples. In general, the more samples are used as exemplars, the higher the weighted F1-score the IL method can achieve. We also find that our method (FastICARL) needs only 5\% budget size to achieves a weighted F1-score of 60-65\% and successfully retain its weighted F1-score even after losing some information by applying quantization up to 8 bits on its exemplars.

\noindent\textbf{Latency:} 
We measure the computational costs of sequentially learning additional classes based on a pre-trained model. The average IL time to run different IL methods is presented in Table~\ref{tab:latency}.
The IL time of FastICARL (32, 16, and 8 bits) ranges 4.96-67.1 seconds on Jetson Nano and 0.83-11.2 seconds on Google Pixel 4 depending on the budget and datasets. FastICARL remarkably reduces the IL time by 18-78\% on Jetson Nano and 37-92\% on Google Pixel 4 compared to ICARL. Note that the training time of ICARL and FastICARL is approximately the same (these results are omitted for brevity). Also, FastICARL (16 and 8 bits) shows substantial improvement in IL time: this indicates that the additional operation of quantizing exemplars does not impose a meaningful burden on the~system.

\begin{figure}[t]
  \centering
  \subfloat[EmotionSense]{
    \includegraphics[width=0.224\textwidth]{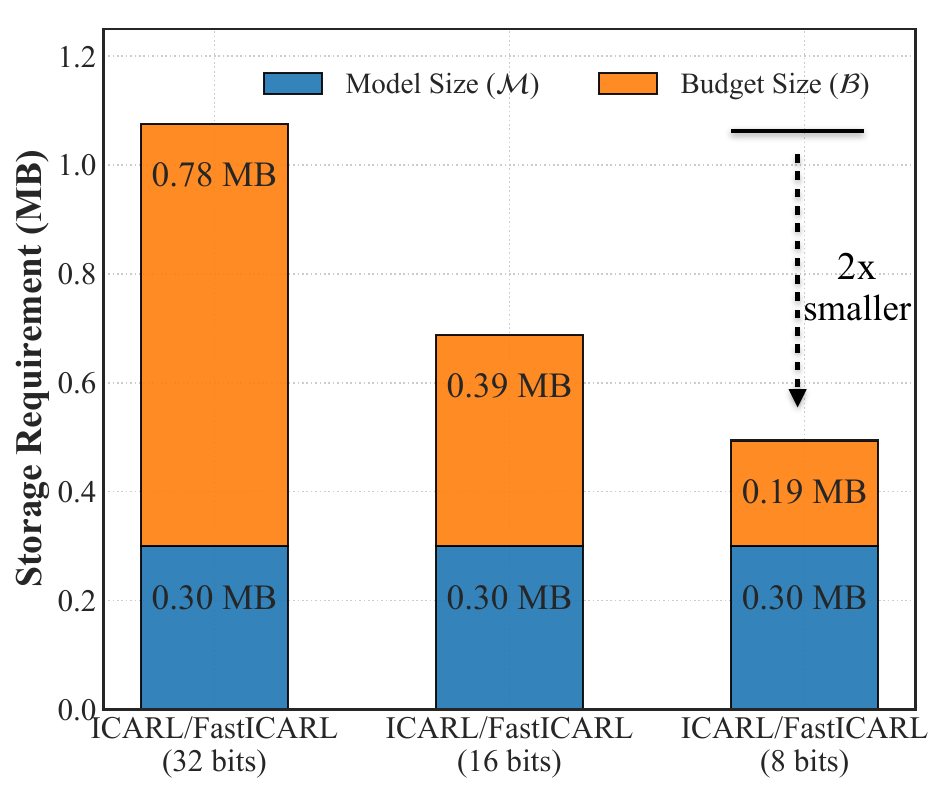}
  \label{fig:emotion}
  }
  \subfloat[UrbanSound8K]{
    \includegraphics[width=0.224\textwidth]{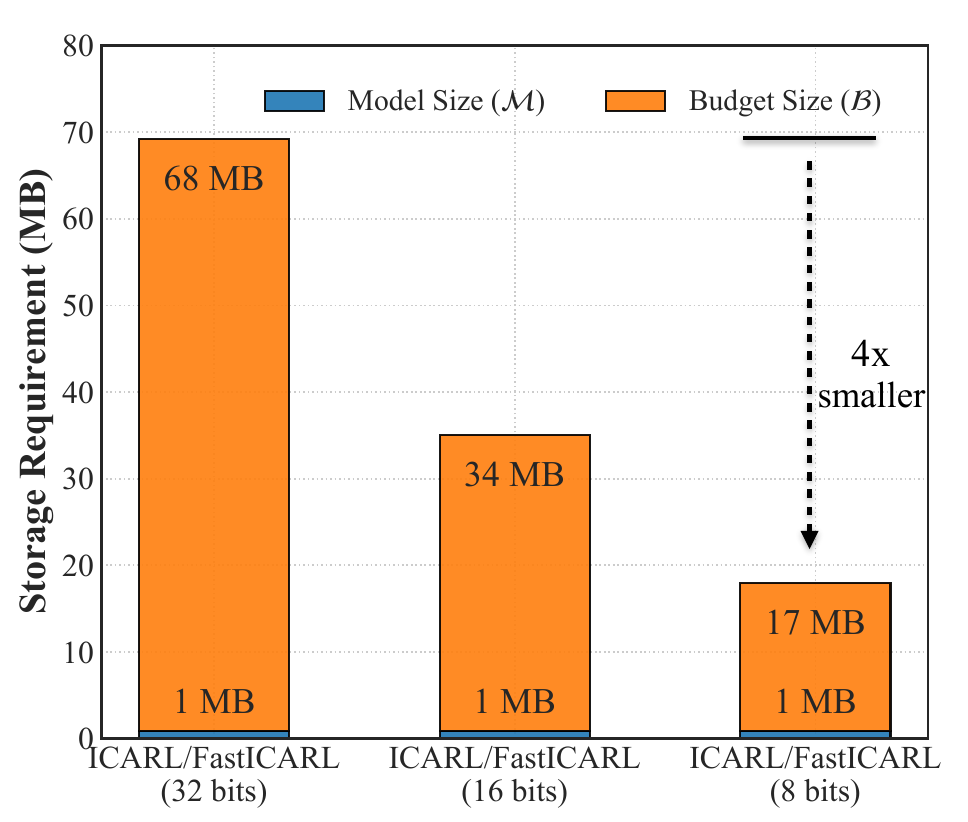}
  \label{fig:urbansound}
  }
  \caption{
  Comparison of the storage requirement ($\mathcal{M}+\mathcal{B}$) for ICARL and FastICARL (32, 16, and 8 bits) based on 20\% budget size in each dataset. 
  }
  \label{fig:storage}
  \vspace{-0.5cm}
\end{figure}

\noindent\textbf{Storage:} 
We now show the storage overhead of the IL method. The size of FastICARL is composed of the model parameter size ($\mathcal{M}$) and budget size ($\mathcal{B}$). As FastICARL relies on stored exemplars, its storage demand is primarily driven by the number of exemplars to be stored, i.e., budget size ($\mathcal{B}$). As shown in Figure~\ref{fig:storage}, FastICARL (8 bits) requires at most 0.49 MB and 18 MB ($\mathcal{M}+\mathcal{B}$) for EmotionSense the UrbanSound8K datasets respectively, decreasing the storage requirement 2 to 4 folds over ICARL (32 bits). Model sizes for EmotionSense and UrbanSound8K datasets are fixed as 0.3 MB and 1 MB, respectively. 

\textit{Based on the results in this section, we have demonstrated that FastICARL enables faster IL by reducing the IL time and storage requirements by applying quantization.}

\section{Conclusions}\label{sec:Conclusions}

In this paper, we developed an end-to-end and on-device IL framework, FastICARL, that enables efficient and accurate IL in mobile sensing applications. We implemented FastICARL on two resource-constrained devices (Jetson Nano and Google Pixel 4) and demonstrated its effectiveness and efficiency. FastICARL decreases the IL time up to 78-92\% by optimizing the exemplar construction procedure and also reduces the storage requirements by 2-4 times by quantizing its exemplars without sacrificing the performance.

There are many interesting directions that deserve further research. First of all, we want to extend our work to enable a higher degree of quantization (such as using 2 or 3 bits) and apply pruning on a model to reduce the model parameters and speed up the training process, which is another bottleneck of the IL. Furthermore, it is worth investigating IL methods on more severely resource-constrained devices such as micro-controller units having meager system resources. 



\section{Acknowledgments}\label{sec:Acknowledgment}
This work is supported by ERC through Project 833296 (EAR) and by a Google Faculty Award. 
We thank Pete Warden for his valuable comments and suggestions.

\bibliographystyle{IEEEtran}

\bibliography{mybib}


\end{document}